# TICAL: Trusted and Integrity-protected Compilation of AppLications

Robert Krahn, Nikson Kanti Paul, Franz Gregor, Le Quoc Do, Andrey Brito, André Martin, Christof Fetzer
Chair for Systems Engineering, Institute of Systems Architecture, Computer Science
Technische Universität Dresden, Germany

*Abstract*—During the past few years, we have witnessed various efforts to provide confidentiality and integrity for applications running in untrusted environments such as public clouds. In most of these approaches, hardware extensions such as Intel SGX, TDX, AMD SEV, etc., are leveraged to provide encryption and integrity protection on the process or VM level. Although all of these approaches increase the trust in the application at runtime, an often-overlooked aspect is the integrity and confidentiality protection at build time, which is equally essential as maliciously injected code during compilation can compromise the entire application and system.

In this paper, we present Tical, a practical framework for trusted compilation that provides integrity protection and confidentiality in the build pipeline from source code to the final executable. Our approach harnesses TEEs as runtime protection but enriches TEEs with file system shielding and an immutable audit log with version history to provide accountability. This way, we can ensure that the compiler chain can only access trusted files and intermediate output, such as object files produced by trusted processes. Our evaluation using micro- and macro-benchmarks shows that Tical can protect the confidentiality and integrity of whole CI/CD pipelines with an acceptable performance overhead.

## I. INTRODUCTION

With the advent of Trusted Execution Environments (TEEs), such as Intel SGX [1], [2], TDX [3] and AMD SEV [4], a lot of privacy sensitive applications can be now safely operated in shared environments such as public clouds. Various TEE (Trusted Execution Environment) technologies provide security guarantees for safe operation. The common feature among these approaches is that they offer confidentiality and integrity protection, which prevents any malicious user with root access on the host machine from performing memory dumps to extract passwords or sensitive data from the application.

While previous literature in the field of TEEs has focused mostly on practical aspects, such as optimizing performance and balancing the size of the TCB with security guarantees, little attention has been given to the trustworthiness of the entire CI/CD pipeline. However, this aspect is just as crucial, since a compromised compiler or build pipeline can potentially allow unauthorized access to sensitive data in the production system.

Protecting the build pipeline is becoming increasingly important as more and more build pipelines are operated in cloud environments these days. Prominent examples are gitlab [5] CI with direct Kubernetes integration or even Cloud-native based approaches such as Tekton [6] where the build pipeline often shares the same public infrastructure as the final application itself.

In this paper, we present a novel practical approach for CI/CD pipelines that protects the whole build pipeline of applications leveraging TEEs to carry out trusted compilation, shielding of the file system and network connections enriched with an immutable audit log. Our approach guarantees that at no point of time, a malicious user can read or modify the source code, i.e., the intellectual property, or tamper with the compiler or build pipeline steps a long the way to the final executable or docker image. Furthermore, using the embedded audit log, all code transformation and build steps will be recorded providing accountability to the entities in charge of development operations, hence, creating a trusted *immutable audit log & version history* comprising all file modifications and transformations over time.

In this paper, we make the following contributions:

(*i*) **Trusted Compilation.** TICAL is leverages SCONE, a TEE runtime which allows to run arbitrary applications in TEEs such as Intel SGX using Docker containers. Using SCONE combined with secure arguments and attestation facility, we can ensure that the compilation process (source code to binary) cannot be manipulated.

(*ii*) **File System Shielding and Audit Log.** We extended SCONE file system shielding layer with integrity protection and audit and version logs by utilizing its system call interception interface. This allows us to trace which file modifications have been performed by which process and entity, reject possible manipulations as well as providing accountability.

(*iii*) **Low overhead performance** We show that our file system-based integrity protection, logging and auditing mechanisms incurs only low overhead making trusted build pipelines feasable Furthermore, we present several performance optimizations to further improve compilation and file access throughput.

The rest of the paper is structured as follows. In Section II, we first provide the background information about the technologies we used in our approach, then we describe the considered threat model. In Section III, we present the overview and the detailed design of TICAL. Next, in Section IV, we describe the implementation details of TICAL. After that, in Section V, we assess the performance of our approach and provide several benchmark results showing the overhead of TICAL when using it in practice. Thereafter, we survey related work in the context of Trusted Execution Environments and audit & version logging in Section VI. We conclude the paper with a summary of contributions and an outlook on future work in Section VII.

## II. BACKGROUND

In this section, we briefly describe the technical building blocks used to build TICAL, our trusted compilation and build pipeline platform.

### A. Shielded Execution

Intel SGX is one of technologies providing trusted execution on process level, hence it ensures that processes are isolated from each other by utilizing a dedicated and cryptographically protected memory region (so called *enclave*) [7]. Enclaves use a contiguous memory region as a block of protected memory borrowed from the Dynamic Random Access Memory (DRAM) as Processor Reserved Memory (PRM). The PRM comprises the Enclave Page Cache (EPC), a set of 4KB memory pages, and enclave meta data which is neither accessible from other applications nor privileged code such as the operating system or the hypervisor. The Memory Encryption Engine (MEE), part of the processor, encrypts and authenticates data for non-PRM memory and protects EPC pages. In contrast to other technologies such as Intel TDX or AMD SEV, we opted for SGX as we want to carry out protection on process-level in build pipelines minizing the TCB.

Futheremore, we leverage the attestation features of Intel SGX which allows a process, i.e., an enclave to prove an enclave's identity to another enclave. During enclave creation, the program data to run the code is first loaded into the enclave memory from the non-protected memory. All subsequent steps that may modify the previously loaded program code are recorded and included in the enclave measurement hash calculation in order to detect potential manipulations. Hence, the measurement hash can later be used to attest an enclave both locally and remotely. We will leverage this menchanisms and use these hashes in our build pipelines to verify if the compiler executables have been modified or not.

The previously described attestation mechanism furtheremore allows us to verify if an application is executed in a legitimate SGX-enabled platform. To perform such a verification also known as remote attestation, SGX transforms a local report into a verifiable quote [7] using the SGX's internal *Quoting Enclave*. The *Quoting Enclave* prepares an SGX attestation signature using an SGX attestation key which is unique to the SGX-hardware. The generated quote can then be delivered to remote platforms for verification [7], [8].

Although the Intel provides software developers with a SDK to run applications in enclaves, the usage of such interfaces is cumbersome as it requires developers to provide glue code for every possible system call an application may use. To overcome this burden, frameworks such as SCONE [9] or Gramine [10] evolved that simplify or even completely eliminate manual steps required to run applications in TEEs using Intel SGX. SCONE achieves this by providing a custom standard c-library which contains glue code as well as all enclave- and system-call handling routines. Additionally, SCONE provides ready to use functionality for attestation and secret provisioning, and an extensible file system shielding interface to provide encryption, integrity protection and audit logging with version history etc.

Since SCONE enables users to run legacy applications in Intel SGX enclaves without modifications through its modified dynamic loader and its extensible shielding layer, we opted to use SCONE as one of the building blocks for TICAL. Futhermore, we opted to use the version control system GIT as it is a mature technology integrate the integrity protection and audit log facility in the shielding layer of SCONE.

### B. Threat Model

For the threat model, we assume a potentially malicious environment in which privileged processes such as the operating system have full control over system calls arguments and their results. In such a compromised system, an attacker can not only modify the system data but can also eavesdrop on system activities. Apart from that, we assume that access to the hardware is strictly regulated and that an adversary cannot mount physical attacks on the otherwise trusted CPU. However, we assume that a malicious privileged software may be installed by e.g. an administrator.

## III. DESIGN

TICAL is designed to transparently support integrity protection and versioning (i.e., immutable and consistent modification history) for legacy applications such as gcc, make etc. in build pipelines. At the high level, our system consists of two core components: ($i$) the build pipeline executables (e.g., gcc, make, node etc.) running inside an enclave using SCONE [9]; and the ($ii$) Configuration and attestation service (CAS) which keeps track of the integrity and history of the modifying content for the applications involved in the build pipeline. TICAL keeps track of any content modification in these applications/processes by maintaining a permanent record history of the changing. Hence, DEVOPS rely on the CAS component [11] to verify the integrity and authenticity of the application. In case the content of the application is encrypted for confidentiality, the secrets are transparently transferred from CAS to the application via a TLS connection after the attestation process.

*1) Versioning support and integrity protection:* Typically, the intermediate as well as final output of build pipelines is stored in the file systems for persistence. To transparently provide integrity protection as well as versioning for build pipelines, TICAL maintains transparently a GIT repository in the local file system. Hence, TICAL introduces an abstraction between the application and the underlying file system of the operating system by intercepting all file-manipulating syscalls executed by the application and transforming these operations into GIT commits. By doing this, all modifications are monitored, audited, and persistently stored into the commit history of the GIT system. This mechanism is integrated through SCONE's libc (i.e., a modified musl C library enabling legacy applications running inside Intel SGX enclave) to support native applications such as gcc etc. without changing their source code. The abstraction ensures that the updated content of the application can be even encrypted before writing down to the underlying file system of the operating system. In addition, this abstraction provides integrity protection for these files by keeping track their metadata and storing them in the CAS component.

*2) Configuration and attestation service:* TICAL relies on an external service component for attestation, secret provisioning and replay protection. Built into the SCONE framework, the Local Attestation Service (LAS) and the external Configuration and Attestation Service (CAS) incorporate the attestation design proposed by Anati et al. [12]. The Configuration and Attestation Service (CAS) attests enclaves, i.e., applications compiled with SCONE and maintains information about the previously attested enclave instances, and provisions configuration to applications. With regard to TICAL, CAS further assists in checking the validity of hash-sums during application startup.

## A. Detailed Design

TICAL's architecture is based on SCONE's file system shielding layer, which protects data and enclave exchanges with the host system at runtime (e.g. file I/O). When reading/writing data from/to the host system, the shielding layer transparently uses cryptographic schemes based on the user's configuration. Outgoing data is hashed and encrypted, while incoming data is decrypted, verified, audited, and versioned.

SCONE's file system shielding layer is made up of the following abstractions: *regions*, *system call handlers*, and *storage providers*. Regions contain information about which storage provider is responsible for a particular folder on the file system. Therefore, a file system-related system call is first mapped to a region and then forwarded to the appropriate system call handler of the storage provider responsible for the region.

By default, SCONE does not encrypt or provide integrity protection for files, i.e. no storage provider is used. However, the user can enable, for instance, the encryption and integrity storage provider, as shown in Figure 1 for the path "/encrypted+authenticated". If enabled, a system call such as a simple write will first go through the encryption storage provider, which encrypts the buffer, and then go through the integrity storage provider, which computes a hash for that buffer or block to detect data manipulations.

It is also possible to use only one of the storage providers, such as the integrity protection provider shown for the path "/authenticated", if the user solely requires integrity protection. To provide integrity protection across whole directories with audit logging and version history, we added another storage provider, GITFS, which uses GIT to record file modifications and provide integrity protection.

*1) The Git File System:* When an application accesses a GITFS-protected region, the storage provider communicates with a remote attestation service over a secure channel to determine whether a fresh empty repository should be initialized or an existing one should be cloned based on the user's configuration. If the user opts to use an existing repository, the git-syscall-handler verifies the repository's integrity using the user-provided hash value. The git-syscall-handler checks the protection policies defined by the storage provider for the .git metadata folder. We use SCONE's integrity storage provider to protect GIT's metadata from outside attacks. All write operations to files that map to regions for which the GITFS storage provider has been enabled are transparently versioned.

Write operations that map to other regions are sent directly to the host system or encrypted if the user opts for the encryption storage provider. Each system call generates a predefined GIT event. GITFS keeps an index with the latest hash value of a root commit in memory to provide integrity protection for a whole directory tree. The git-syscall-handler can periodically send this hash value to a remote server using a secure communication channel to protect the underlying repository from rollback attacks.

As mentioned previously, we use GIT for versioning as it provides integrity protection out of the box that ($i$) spans whole directory trees and ($ii$) version history by chaining the commit objects (hash chain). We utilize some of SCONE's basic integrity mechanisms to provide a fully protected stack, since the metadata of a GIT repository itself is not protected against data manipulation attacks. SCONE provides only file-level integrity protection, unlike GIT which protects entire directory trees with version history.

To keep the overhead for integrity protection small, we identified the minimal set of metadata resources in a GIT repository that require protection. For the minimal setup, we grouped GIT's metadata into two regions: The *revision log* and the *database*.

The *database* (.objects file) stores all the file data and their modifications, while the *revision log* contains the hashes (integrity protection) and pointers to the file data in the order of commits. Protecting the revision log ensures the integrity of the database. For the minimal setup, GIT's metadata is divided into two regions: an *unprotected* region for the *database* and an *authenticated* region (integrity protection without encryption) for the *revision log*.

A commit in GIT can occur independently of file modifications, which creates a vulnerability. A malicious user can modify the data between the write and the commit. Therefore, we introduce secure buffers where all file modifications are first written to a buffer and only flushed out to the next lower storage provider or host system once the commit has been executed successfully.

## IV. IMPLEMENTATION

After laying out the architecture and design of TICAL in the previous section, we will now provide more implementation

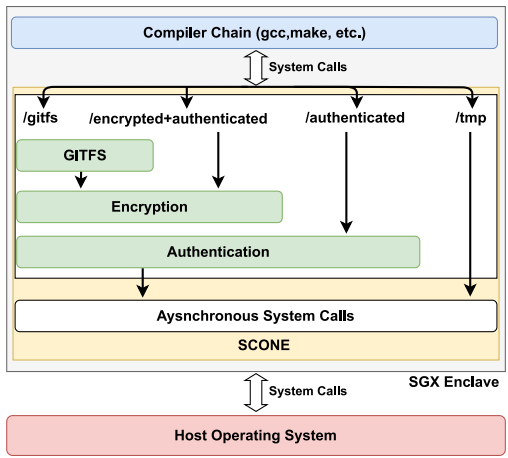

Fig. 1: TICAL architecture for versioning support.

| syscalls | LIBGIT functions |
|---|---|
| `open, openat` | git_verify_buffer |
| `write, writev, pwritev` | 1. git_stage_buffer<br>2. git_commit |
| `read, readv, preadv` | (Use buffer) |
| `linkat, rename, renameat` | 1. git_rename<br>2. git_commit |
| `unlink, unlinkat` | 1. git_remove<br>2. git_commit |
| `clone, fork, execve` | 1. git_commit<br>2. git_push |

TABLE I: One-to-one mapper: System calls to LIBGIT functions

details of the system.

TICAL is tightly integrated into SCONE, a TEE runtime which provides shielded execution for legacy applications. The integration is done by utilizing SCONE's extensible layered file system interface which is based on system calls interception using a customized *libc* based on *muslc* [13].

For integrity protection, auditing and versioning, we use LIBGIT [14], a library that provides the functionality to operate on GIT repositories typically used for source code repositories. The library provides an Application Programming Interface (API) which is equivalent to the Command Line Interface (CLI).

### A. GIT Integration

In order to fully integrate LIBGIT as a storage provider in SCONE's shielding layer, we had to adapt LIBGIT as follows: First, we removed any usage of the Linux utility `realpath` which LIBGIT uses to resolve relative paths and absolute pathnames of symbolic links [15]. Since symbolic links are not used in the shielding layer and not supported by SCONE, we removed this functionality from the LIBGIT source code. Second, we moved the *staging* area to an in-memory file system inside of an enclave as we perform staging and committing (git add & git commit) as an atomic step in the protected memory of an enclave.

Besides using LIBGIT to record the history for file changes, we use the library also to verify the integrity of the contents of a repository (using its hash), upon ($i$) cloning of an existing repository and ($ii$) whenever file data is being read through a read system call.

### B. GITFS Implementation

As mentioned previously, LIBGIT is integrated as a storage provider in SCONE's shielding layer. This has several advantages: First, it provides automated and transparent versioning of sensitive information. Second, the file contents, and the state of whole directory trees are integrity protected. For our GITFS storage provider, we implemented our own syscalls-handler named GITFS-HANDLER.

However, implementing a syscalls-handler is not a trivial task as the Linux Kernel provides hundreds of syscalls [16] for privileged operations where many of them provide the same functionality but have different signatures. In general, applications use system calls to communicate with the Linux Kernel to execute privileged tasks, such as copying a memory buffer or reading data from the storage device. In the case of a file system as in GITFS, we would have to intercept the `read`, `readv`, `preadv`, `preadv2`, and `pread64` system call to provide integrity protection when reading a file from the storage system.

Fortunately, SCONE's syscalls-handler allows the grouping and mapping of system calls of the same type to a single system call. Thus, we only need to handle the `preadv` system call in the GITFS storage provider in order to cover all read operations.

In order to achieve optimal performance, we also maintain some extra internal data structures in the GITFS storage provider besides using existing LIBGIT's functionality. For example, in order to allow fast path traversals happening during the staging phase of git prior to a history commit, we keep the original file descriptor, access modes, creation data and flags as well as a buffer with the full filename in memory.

### C. In-Enclave Buffer

To ensure strong integrity protection, we must prevent file contents from being manipulated at any time. However, git does not allow the direct retrieval of file contents from its repository database, i.e., the *git* directory. So, to access the file contents, one must first checkout the file and retrieve it from the working copy of the repository. Similarly, changes to the file are written to the working copy and then staged before being integrated into the repository's database. This creates a window of vulnerability for malicious users to modify files in the working copy undetected.

To prevent such manipulations, we introduce the concept of an *In-Enclave Buffer*, a memory region inside an enclave that cannot be tampered with. The secure buffer is used to write changes to the file instead of writing them to the working directory. The buffer functions as an in-memory file system, and other operations can be performed on it, such as computing a hash, preventing data manipulations in between. When reading files, they are first read from disk into the secure buffer, and integrity is checked using LIBGIT before being handed over to the application as a user buffer.

*D. Metadata Regions*

In a nutshell, a git repository consists of two different entities: the `.git/objects` directory which contains the file contents and history of all files versioned in the repository, and the *refs* and *logs* files which hold the metadata for the versioned files as well as the history. Using the metadata, it is possible to detect integrity violation either on the checked'out files in the working directory or the history stored in the `.git/objects` directory. Hence, the most vulnerable part of a git repository is the metadata itself represented by the *refs* and *logs* files stored in the `.git` directory.

In order to provide a fully integrity protected system, we therefore defined two distinct regions with different properties: Region A comprises the *refs* and *logs* in the `.git` directory while region B contains the `.git/objects` directory and the remaining files of the `.git`.

We protect region A using the authenticated kernel region SCONE provides out of the box to detect integrity violations. We leave region B unprotected. We do not protect region B as ($i$) integrity protection incurs a certain overhead as we will show in the evaluation section of this paper and ($ii$) since manipulations of the `.git/objects` can be easily detected using the metadata of region A.

*E. System-calls and* GIT *functions Mapper*

Implementing a storage provider in SCONE is relatively straightforward as the intercepted file systems operations such as opening and closing, reading and writing a file must be solely replaced with a custom implementation carrying out a different functionality. Table I summarizes the actions taken by our GITFS storage provider upon execution of a specific file system operation.

For example, an `unlink` or `unlinkat` system call usually deletes a file from the file system. We intercept the original call and execute the following action instead: We first update the in-memory stage file in conjunction with a git commit to track that the file is deleted. We then check the file in the cached stage using `scone_git_available_in_index`, and remove it from the cached stage `scone_git_remove`. Finally, the system call handler forwards the system call to the kernel and creates a git commit upon a successful execution.

Looking at the mapping of Table I reveals that a commit is executed with every single write to a file. This provides a very fine granular history as even a write of a few bytes create a new history entry. However, in case the user opts for a more coarse grain history, it is possible to record file modifications only upon the time the application closes a file.

We therefore introduce two modes for the file system call mapper: (A) Fine-grained and (B) Coarse-grained. In Fine-grained mode, a new history entry is created with every single `write` call while in coarse-grained mode only upon a `clone`, `fork` and `sync`.

Table II depicts the intercepted system calls for the different modes.

| Mapper Mode | syscalls |
|---|---|
| Coarse-grained | `clone, fork, vfork, execve, fsync, exit` |
| Fine-grained | `open, openat, creat`<br>`write, writev, pwritev`<br>`unlink, unlinkat`<br>`linkat, rename, renameat` |

TABLE II: GITFS mapper mode

*F. Read and Write Application Calls*

This section briefly describes the workflow when GITFS encounters a `read` or `write` system call. If an application opens a file in read mode, the GITFS storage provider receives an open system call from the SCONE syscall-handler. The GITFS file open handler checks if the file exists in the underlying storage system and opens the requested file. After opening the file, it reads the contents into the secure buffer and calculates the hash value of the previously read file contents. Then, GITFS verifies the calculated hash against the hash value provided by LIBGIT through the stored metadata. If the hashes match, the file data has not been manipulated and can be safely passed on to the requesting application. However, in case the verification fails, the system aborts the application's execution. Note that the system will always abort if any file in the GITFS has a mismatch with the hashes recorded in the metadata of the repository. This approach also applies to file manipulations other than the currently processed file, as its addition to the staging area will ultimately result in the staging area's hash mismatch.

In addition to the individual hashes for files, the repository also maintains a root hash of the hash chain in memory, which is only updated by the system during the operation and if the file operations do not trigger any integrity violation.

As a result, GITFS can detect any unwanted or malicious data modification from the tip of the in-memory stage's history, i.e., the root hash. This measure protects against rollback attacks if the in-memory root hash value is continuously transferred to a trusted service.

In a similar way as with `read` system calls, `writes` are intercepted using the GITFS storage provider in the following way: The SCONE syscall-handler maps the different write calls to the `pwritev` system call first in order to simplify the storage provider implementation. Hence, our GITFS storage provider contains only the implementation for the `pwritev` system call to handle all write operations to files. If the GITFS storage provider encounters a write call, the SCONE syscalls-handler processes and forwards the system call to the appropriate GITFS-HANDLER. The GITFS-HANDLER assigns the operation to the GITFS system-call-mapper handler (§IV-E) in the GITFS storage provider. The GITFS storage provider then generates the hash value of file data to be written and updates the in-memory staging file in the GITFS using the secure buffer (§IV-C). After updating the secure buffer, GITFS executes git staging (`scone_git_stage_buffer`) of that buffer to update the in-memory *staging area*. As a result, the in-memory staging area already holds the data to be written in the GITFS file system. However, neither the root hash value is updated, the updated in-memory stage is not saved in the file system, nor a commit has been created. We delay those operations because GITFS first dispatches the system call to the host operating system. As soon as the operation successfully returns, GITFS creates a commit using the in-memory staging area, followed by an update of the in-memory root hash value, which is saved in the in-memory staging area and the metadata region. Note that the SCONE file system shield protects the metadata of the GITFS repository. Data hash values and git objects' hash values are calculated before forwarding the system call to the untrusted operating system to ensure their integrity. Any modification in the content by the compromised operating system can be detected in subsequent read or write operations. System log commit messages include essential metadata such as enclave-id, process-id, system call number, among others.

| GITFS Variant | integrity | encrypted | commit on write | git payload prot. | in memory |
|---|---|---|---|---|---|
| Nopr-coarse-disk | | | | ✓ | |
| Nopr-fine-disk | | | ✓ | ✓ | |
| Inte-fine-disk | ✓ | | ✓ | ✓ | |
| Encr-fine-disk | ✓ | ✓ | ✓ | ✓ | |
| Encr-coarse-disk | ✓ | ✓ | | ✓ | |
| Inte-coarse-disk | ✓ | | | | |
| Inte-fine-mem | ✓ | | ✓ | ✓ | ✓ |

TABLE III: Evaluation variants with protection and optimization combinations for evaluation

## V. EVALUATION

In this section, we present the evaluation of TICAL using micro-benchmarks and a real-world case-study.

### A. Experiment Setup

We executed the performance evaluation experiments on a system with an Intel Xeon E3-1280 v6 3.90GHz processor, 64 GB of main memory, and a 500 GB SATA-based SSD. The host machines run Ubuntu 20.04 with a 64-bit Linux kernel (5.4.0). The Gcc [17] and glibc [18] versions are 9.4 and 2.31. All experiments are conducted inside Docker containers (version 20.10.19) using Ubuntu 20.04.

### B. Performance Benchmark

The system performance is evaluated with multiple variants of GITFS that differ in protection modes and optimizations. The combinations are presented in Table III.

For example, *Encr-coarse-disk* denotes a variant that protects the integrity of persisted application data and GITFS meta-data *.git/*, as well as ensuring that all data is encrypted on disk. Additionally, this variant creates GIT commits only upon rare system calls (see Table II). *Inte-fine-mem* denotes a variant that protects the integrity of persisted application data and GITFS meta-data *.git/* but keeps all data in memory until *git push* is called internally, e.g. when IOZone [19] closes the file.

We use SCONE to protect the GIT meta-data directory and its subdirectories through encryption or by applying integrity protection. As explained in (§III-A1) it is not necessary to apply integrity protection to the *.git/objects* directory to achieve integrity protection of the working area (application data). For this reason, we exclude the *.git/objects* directory from the integrity protection in the variants *Inte-fine-disk* and *Inte-coarse-disk*. For the variants *Encr-fine-disk* and *Encr-coarse-disk*, we encrypt the *.git/objects* directory to ensure data privacy.

We use IOZone [19], a file system benchmark tool, for measuring the throughput of the file system layer in TICAL. IOZone executes various file system operations, such as `read`, `write`, `fread`, `fwrite` and `mmap`, in a file system to measure the system I/O performance. Users can define a total file size and record size for the tests. The record size defines the argument length per system operation, whereas the file size defines the total data to be written to disk. For example, for a measurement with a file size of 2 MB and 16 KB record size, IOZone writes 16 KB data per write operation until 2 MB have been written to disk. We compiled IOZone using SCONE and the different variants of GITFS (Table III).

### C. Performance Measurements

During the system's write performance tests, we evaluate different file sizes against a record size of 16 KB. Figure 2 shows the write-throughput for the depicted configurations whereas Figure 2 the measurements for the *rewrite* command. Variants that make use of the *Fine-Grained* mode show a steep performance decline and very low overall throughput. In Figure 2 *Inte-fine-disk* achieves only 231.5 KB/s for files of 32 KB in size, while the throughput of *Encr-fine-disk* is 176.5 KB/s for a file size of 32 KB/s.

After examining TICAL's use of GIT within the protection system, we discovered that the reason for this overhead is the protection layer of GITFS itself. Due to the fact that GITFS recalculates the complete file hash for a partial rewrite of data, generates commit meta-data and accesses additional meta-data on disk, we find that the performance of the fine-grained *Fine-Grained* variants can be improved by executing *commits* less frequently. Therefore, variants that use the *Coarse-Grained* mode issue a commit only on rare system calls, e.g. *close*. We also discovered that the *Coarse-Grained* mode allows GITFS to benefit from increasing file sizes as commits are issued less frequently with increasing file sizes. While *Encr-coarse-disk* achieves only a write throughput of 283.1 KB/s for 32 KB files, the throughput increases to 37 MB/s for files of 64 MB size.

For the read performance evaluation of TICAL we use the same datasets and variants as for the write performance evaluation. In Figure 3, we present the results for reading data during the IOZone benchmark.

Note that GITFS benefits from the internal secure buffer (§III-A1) in the read operations. It serves files from the internal memory instead of forwarding the system calls to the host operating system. Additionally, GITFS verifies data at the time of opening and a succeeding read uses the verified cached data from the secure buffer. Nevertheless, SCONE verifies data integrity for each read operation.

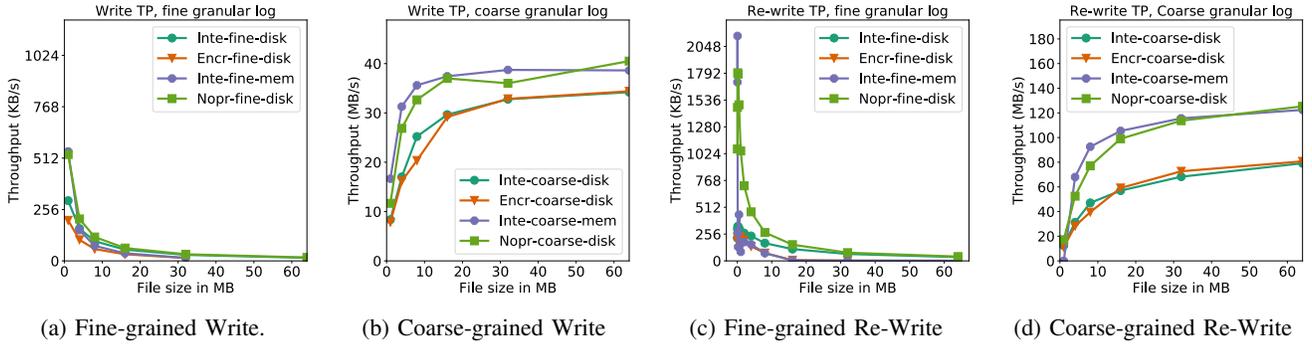

(a) Fine-grained Write.  (b) Coarse-grained Write  (c) Fine-grained Re-Write  (d) Coarse-grained Re-Write

Fig. 2: IOZone Write / Re-Write Throughput for Fine-Grained and Coarse-Grained Variants.

Similar as in the previous results, Figure 3 shows a stark performance reduction for variants that use the *Fine-Grained* mode.

Given the presented results, we further analyzed the TICAL for bottlenecks and implementation deficits. Our investigation revealed that extensive use of system calls and data amplification significantly reduce the performance of GITFS.

For example, a single `write` operation by IOZone that results in a git-commit creates up to 8 accesses on disk: Per commit process to retain the change history made by an application, GITFS creates three object directories in the GIT database directory, writes three git objects, updates two logs and two references, and finally updates the stage file. Also, GIT generates intermediate temporary files before each git object file is created. Apart from that, we found that GIT uses (*fstat*) system calls prior to any operation for file status checking. Furthermore, the Git *push* operation dramatically reduces the throughput of all variants. Figure 4 shows a comparison of two variants with *push* and without *push*. Without executing the *push* operation *Encr-coarse-disk* achieves a write rate of 96 MB/s for 64 MB large files.

Additionally, GITFS updates an internal security metadata file during each write in the protected regions for the freshness of check-sums. For this reason, GITFS executes 61% additional `open` and `close` system calls for the protection metadata. Also, it performs 70% more system calls for memory allocation compared to non-Scone applications.

## VI. RELATED WORK

The body of related work can be largely broken up into two fields: (*i*) Approaches that focus solely on the integrity protection mechanisms and include the host operating system in their trust model, and (*ii*) approaches that take a more holistic view considering new technologies such as Intel SGX broadening the overall threat model.

In the first field, the number of publications covered mechanisms for integrity protection of persisted data and how to provide secure application logging using secure file systems [20]–[23]. For example, for Ext3Cow [20], [21] Burns et al. describe techniques on how to provide secure audits and compute MACs incrementally while Oprea et al. [22] focus on aspects such as detecting integrity violations using Merkle-Trees. TICAL goes beyond those works as we assume these techniques as state of the art and utilize them either directly or indirectly. For example, we use the mature LIBGIT library implementation that uses Merkle-Trees internally to detect integrity violations etc.

With the recent availability of trusted execution environments such as Intel SGX, the focus has shifted as researchers begun to exclude the host operating system from their trust model. A number of publications evolved in this second field as the new trust model requires novel solutions which rely less on services offered by the operating system but at cloud computing level as well as distributed systems. For example, authors in ROTE [24] Matetic et al. argue that Intel SGX is susceptible to rollback attacks which can break the integrity protection of persisted application state. ROTE builds upon a distributed integrity protection scheme where multiple machines are used to infer the current value of a counter that is used for integrity protection. In contrast to ROTE, TICAL ensures integrity protection across reboots by propagation the root hash continuously to a trusted service entity. The implementation of the trusted service may as well follow a distributed approach but it is not the scope of this work.

Authors in SGX-Log Karande et al. propose secure system logs through a combination of security mechanisms of Intel SGX for sealing persisted data and hash-chaining for efficient integrity verification of logs [25]. TICAL uses similar techniques however goes beyond this approach as it provides protection for arbitrary application data on file systems rather than solely system logs. Furthermore, TICAL is completely transparent and application agnostic, hence does not require any source code or application modification and uses well established algorithms for recording history through LIBGIT.

An approach that focuses on discovering service integrity violations of network-based services is present by the authors of libSEAL. Aublin et al. describe a tool to provide an auditing library for internet-based services that captures client-server interactions and securely logs them prior to relaying the messages to their destinations. Although TICAL is similar as it also provides an audit log that can be conveniently inspected as in libSEAL, TICAL provides auditing on file system level rather than network traffic/interaction level.

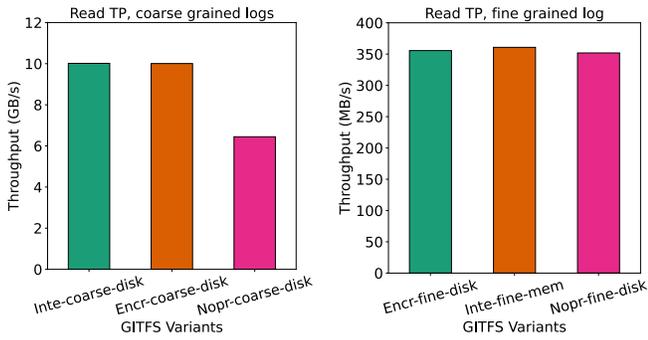
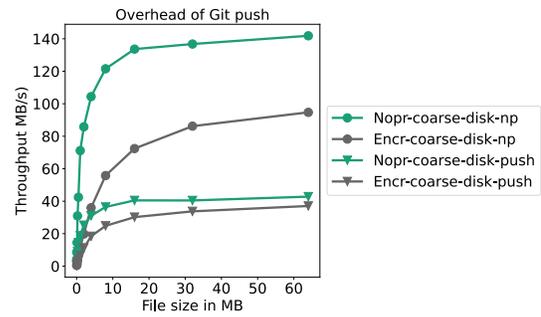

Fig. 3: IOZone Read Throughput for Fine-Grained and Coarse-Grained Variants. File Size = 64 MB.

Fig. 4: IOZone Write Throughput with Git Push enabled and disabled.

## VII. CONCLUSION AND FUTURE WORK

In this paper, we presented TICAL, an auditing log and integrity protection layer for file systems using Intel SGX. TICAL provides users with a trustworthy, immutable audit log & version history of file modifications that can be conveniently inspected using industry standard tools such as GIT clients. In addition to the log, TICAL automatically detects integrity violations and aborts applications to ensure that at no point in time applications operate on manipulated data.

TICAL is completely transparent as it is based on SCONE and tightly integrated into the standard libc library. Hence users neither have to perform any source code modifications nor re-compilations of their applications.

We evaluated the performance of TICAL with regards to the overhead it introduces in general as with its different configuration variants that allows recording of file modification history at different granularity levels as well as protection guarantees.